\documentclass[manuscript]{aastex61}

\newcommand\aastex{AAS\TeX}

\newcommand{\R}{\mathbb{R}}

\usepackage{subfigure}

\submitjournal{ApJ}

\shorttitle{\aastex\ A hybrid machine learning approach to flare prediction}
\shortauthors{Benvenuto et al.}

\begin{document}

\title{A hybrid supervised/unsupervised machine learning approach to solar flare prediction}

\correspondingauthor{Michele Piana}
\email{piana@dima.unige.it}

\author{Federico Benvenuto}
\affil{Dipartimento di Matematica Universit\`a di Genova, via Dodecaneso 35 16146 Genova, Italy}


\author{Michele Piana}
\affil{Dipartimento di Matematica Universit\`a di Genova and CNR - SPIN Genova, via Dodecaneso 35 16146 Genova, Italy}

\author{Cristina Campi}
\affiliation{CNR - SPIN Genova, via Dodecaneso 33 16146 Genova, Italy}

\author{Anna Maria Massone}
\affiliation{CNR - SPIN Genova, via Dodecaneso 33 16146 Genova, Italy}



\begin{abstract}
We introduce a hybrid approach to solar flare prediction, whereby a supervised regularization method is used to realize feature importance and an unsupervised clustering method is used to realize the binary flare/no-flare decision. The approach is validated against NOAA SWPC data.
\end{abstract}




\section{Introduction} \label{sec:intro}

Solar flares are the most energetic events in the solar system. Over a typical duration of $\sim$(10 - 1000)~s, they can release up to $10^{32}$~erg of energy --- stored in stressed active region magnetic fields --- into directed mass motions, heating, and acceleration of supra-thermal charged particles, including electrons, protons and heavier ions \citep{kontar_et_al}.  Solar flares, together with Coronal Mass Ejections (CMEs), are the main drivers of space weather at Earth and can sometimes even significantly affect Earth- and space-based technology systems like power grids, flight navigation, satellite communications. Predicting solar flares requires, first of all, the determination of parameters such as properties of sunspot groups or of the coronal magnetic field configuration, that are thought to be important for the understanding of fundamental processes in solar plasma physics. Second, at a more technological level, these parameters are used as input values for algorithms that realize predictions providing, for example (but not exclusively), a binary flare/no-flare outcome \citep{bletal12,wh04,gaetal02}. 

Most recent flare prediction algorithms belong to the machine learning framework \citep{boco15,coqa09,lietal07,yuetal09,yuetal10}. In this setting, data properties utilized for prediction are named {\em{features}}. In the case of {\em{supervised}} learning, a set of historical data is at disposal where features are tagged by means of {\em{labels}} representing the observation outcome, and the prediction task consists in determining the label associated to the incoming features' set. On the other hand, {\em{unsupervised}} methods do not use any training set and data are clustered in different groups according to similarity criteria involving data features.

A crucial aspect of flare prediction, characterized by notable physical implications, is to provide hints on which data features mostly correlate with the labels. In statistical learning theory this practice is known as {\em{feature selection}} although applications often refer to it as {\em{feature importance}}, which better points out the fact that at the end of the process features are ranked according to their importance in the prediction task (in the following we will use the two terms equivalently). Feature selection can be realized by using advanced implementations of standard neural network approaches \citep{oletal04,ga91} or by means of regularization methods. This second approach aims to optimize a functional made of two terms: the discrepancy term measures the distance between prediction and data in the training set, while the penalty term imposes a constraint on the number of features that significantly contribute to the prediction itself. Two examples of regularization methods for feature selection are LASSO \citep{tibshirani_regression_1996} and $l1$-penalized logit ($l1$-logit in the following) \citep{wu_genome-wide_2009}. Both approaches utilize an $l1$-norm penalty term to reduce the complexity; on the other hand, LASSO measures the discrepancy assuming that the noise on the data is Gaussian, while $l1$-logit relies on a maximum likelihood procedure in which the probability function to maximize is the binomial distribution. In the framework of flare prediction, each one of these two methods presents a specific limitation. In fact, LASSO is intrinsically a regression method and therefore it is not naturally appropriate for applications like flare prediction that may require a binary yes/no response. On the other hand, $l1$-logit is a classification method, but it predicts the binary condition by applying a fixed threshold on the flare occurrence probability, i.e., the applied threshold is the same whatever the dataset used for training is.

The present paper introduces a novel approach to flare prediction with feature importance whose aim is to overtake both previous limitations. The perspective of such approach is hybrid and rather general: first, a regularization method for regression is applied to the training set with an $l1$ penalty term that promotes sparsity, thus realizing feature importance (more specifically, this regularization step reconstructs the vector of weights, with which each feature contributes to the prediction in the training set). Then, the set of the real values obtained by multiplying the weights times the feature values in the training set is automatically clustered in two classes by means of a clustering technique. Clustering is an unsupervised learning approach that organizes a set of samples into meaningful clusters based on data similarity. Data partition is obtained through the minimization of a cost function involving distances between data and cluster prototypes. Optimal partitions are obtained through iterative optimization: starting from a random initial partition, samples are moved from one cluster to another until no further improvement in the cost function optimization is noticed. Therefore, in the second step of the hybrid approach, clustering performs an automatic thresholding, which depends on the historical set used for the training phase and is, therefore, intrinsically data dependent. The resulting algorithm presents several advantages with respect to standard one-step approaches: it selects the most significant features, since, in the first step, it relies on a regularization technique that promotes sparsity; it is a classification method, since at the end it produces two clusters, each one corresponding to a specific outcome of the prediction;  it performs classification in a flexible, data-adaptive way, which makes it significantly efficient in providing good performances with respect to standard skill scores. The hybrid approach in this paper utilized LASSO in the regularization step and Fuzzy C-means \citep{bezdek_pattern_1981} to cluster the LASSO outcome, although other feature selection and clustering algorithms can be applied.

In order to corroborate the effectiveness of this hybrid approach we utilized a set of data from the National Oceanic and Atmospheric Administration (NOAA) Space Weather Prediction Center (SWPC) and compared our results with the ones provided by $l1$-logit, as far as both the classification and feature importance abilities are concerned, and with some of the most used machine learning approaches in flare forecasting, as far as just the prediction effectiveness is concerned.

The plan of the paper is as follows. Section 2 illustrates the kind of data, prediction algorithms will deal with. Section 3 introduces our hybrid approach for flare prediction with feature importance. Section 4 applies the hybrid approach to the set of SWPC data described in Section 2 and compares its performances to the ones obtained by $l1$-logit and by other machine learning methods. Our conclusions are offered in Section 5.

\section{Categorical data}
Solar Active Regions (ARs) are classified according to magnetic field complexity indicators. For example, ARs tracked by the National Oceanic and Atmospheric Administration (NOAA) Space Weather Prediction Center (SWPC) are typically classified by using the following $5$ indicators (features): the area, the McIntosh indices \citep{mcintosh_classification_1990}, and the Mount Wilson index \citep{hale_magnetic_1919}. The area index is computed in fractions (millionths) of a solar hemisphere. The McIntosh scheme uses white light emissions to represent sunspot structure and is composed by three independent variables: the {\em{Zurich class}} $Z$ of leading/trailing spot size and separation, which may assume $7$ categorical values; the {\em{penumbral class}} $p$ of primary spot regularity, which may assume $6$ categorical values; the {\em{compactness class}} $c$ of internal spot distribution, which may assume $4$ categorical values. Finally, the Mount Wilson scheme groups sunspots into classes based on the complexity of magnetic flux distribution in associated active regions, according to rules set by the Mount Wilson Observatory in California; this feature may assume $8$ categorical data.

In order to apply machine learning algorithms, either supervised or unsupervised, we need to transform the categorical information contained in the above sunspot classifications (specifically, the McIntosh and Mount Wilson indices) into numerical data. This can be done by either transforming the categorical variables into {\em{dummy variables}} \citep{hardy_regression_1993} or computing occurrence frequencies in a historical database. In this paper we used this second approach, which preserves the dimension of the space where to perform the data analysis.  Specifically, we have considered the SWPC database covering the December 1988 to June 1996 time range and we have computed the frequency with which a sunspot classified by a specific value of a fixed indicator produces a flare greater than a given class. Anyhow, we have verified that the use of the dummy variables does not improve the effectiveness of the prediction for all methods considered in this paper. On the other hand, we are also aware that the use of frequencies requires the availability  of a labelled dataset whose content, in principle, may increase while new data are at disposal.

More formally, and focusing on the specific case of the value $A$ for the Zurich class in the McIntosh classification, we denoted by $N_{Z=A}^{(C1\rightarrow C9)}$, $N_{Z=A}^{(M1\rightarrow M9)}$, and $N_{Z=A}^{(\ge X1)}$ the occurrences of flaring events of class $C$, $M$ and $X$, respectively, and computed the frequencies associated to flaring events of class greater or equal to a specific class as
\begin{equation}\label{eq:fc1} 
f_{Z=A}^{(\ge C1)}  = \frac{N_{Z=A}^{(C1\rightarrow C9)} + N_{Z=A}^{(M1\rightarrow M9)} + N_{Z=A}^{(\ge X1)}}{\#~A~occurrences}
\end{equation}
(with  the corresponding no-flare-event frequency defined as $f_{Z=A}^{(no flare)}:= 1 - f_{Z=A}^{(\ge C1)}$);
\begin{equation}\label{eq:fm1}
f_{Z=A}^{(\ge M1)}  = \frac{N_{Z=A}^{(M1\rightarrow M9)} + N_{Z=A}^{(\ge X1)}}{\#~A~occurrences}
\end{equation}
(with  the corresponding no-flare-event frequency defined as $f_{Z=A}^{(no flare)}:= 1 - f_{Z=A}^{(\ge M1)}$);
\begin{equation}\label{eq:fx1}
f_{Z=A}^{(\ge X1)}  = \frac{N_{Z=A}^{(\ge X1)}}{\#~A~occurrences}
\end{equation}
(with  the corresponding no-flare-event frequency defined as $f_{Z=A}^{(no flare)}:= 1 - f_{Z=A}^{(\ge X1)}$). Similar formulas can be written for each one of the other categorical predictors. 

We finally notice that the same dataset used for computing these frequencies has been used as training set for the supervised machine learning algorithms utilized in the following. On the other hand, the database of SWPC indicators covering the time range between August 1996 and December 2010 has been used as test set for both the supervised and unsupervised machine learning methods.

\section{The hybrid approach}\label{pred_alg}

We denote with $X$ the matrix with dimension $N$ (number of active regions) $\times$ $F$ (number of features) whose columns contain the feature values for each specific active region in the training set; $\beta$ is the $F \times 1$ vector containing the $F$ model parameters to determine and $y$ is the $N \times 1$ data vector used in the training set and made of $0$ and $1$ values. $l1$-logit has been designed 'ad hoc' to perform classification with feature importance \citep{wu_genome-wide_2009}. This is a constrained maximum-likelihood method that allows the estimation of the model parameters while best-fitting the data. The logit parameter estimation method solves the minimum problem
\begin{equation}\label{logit-3}
{\hat{\beta}} = \arg\min_{\beta} [-\sum_{i=1}^N \log (1+e^{-y_i(X_i \cdot \beta +c)})]~,
\end{equation}
i.e., it searches for the maximum likelihood of the model parameter vector $\beta$ when one assumes that each component $y_i$ of the vector $y$ is the realization of a random variable described by the Bernoulli distribution. In equation (\ref{logit-3}) $X_i$ is the $i$-th row of matrix $X$ and $c$ is a positive constant. In order to realize feature selection, $l1$-logit adds the condition that $\| \beta \|_1$ is small, which mathematically points out the few parameters that most significantly contribute to the classification. When a new active region $x$ is at disposal ($x$ is a vector of $F$ components), then ${\hat{y}}=\sum_{k=1}^F x_k {\hat{\beta}}_k$ is computed and its sign denotes the outcome of the prediction. This implies that the classification threshold here is fixed and equal to zero independently of the dataset used for training.


We now introduce an approach to flare prediction with feature selection which, differently than $l1$-logit, is hybrid and data-dependent. The first step of this two-step approach utilizes LASSO \citep{tibshirani_regression_1996} to perform feature selection. Specifically, we look for the solution of the minimum problem
\begin{equation}\label{hybrid-1}
{\hat{\beta}} = \arg\min_{\beta} (\|y - X \beta \|_2^2 + \lambda \|\beta \|_1)~,
\end{equation}
where the regularization parameter $\lambda$ is optimized by means of a Cross Validation procedure \citep{stetal74}. Then, in the second step, we apply a clustering method for partitioning the output of ${\hat{y}}=X{\hat{\beta}}$. In a classical clustering approach like Hard C-Means (HCM) \citep{jaetal99}, each sample may belong to a unique cluster, while in a fuzzy clustering formulation a different degree of membership is assigned to each sample with respect to each cluster, which implies a much higher flexibility in accounting for data characteristics. Therefore, in the second step of our hybrid approach, we used Fuzzy C-Means (FCM) \citep{bezdek_pattern_1981}, which is the fuzzy extension of HCM. In this framework, the FCM functional is given by
\begin{equation}\label{FCM-1}
J_m({\hat{y}},{\hat{z}},U) = \sum_{k=1}^N \sum_{j=1}^C (u_{jk})^m d_{jk}^{2}~,
\end{equation}
where ${\hat{z}}=\{{\hat{z}}_j | {\hat{z}}_j \in \R~,~j=1,\ldots,C\}$ is the set of the $C$ centroids of the clusters, the component $u_{jk} \in [0,1]$ of the $C \times N$ matrix $U$ represents the membership of the $k$-th sample to the $j$-th cluster, $d_{jk}$ is the distance between the $j$-th centroid and the $k$-th sample, and $m$ is the fuzzifier parameter. The FCM optimization problem is the one to (iteratively) determine the components of the matrix $U$ and of the vectors ${\hat{z}}$ given the components of the vector ${\hat{y}}$.

\section{Application to SWPC data and analysis of results}

In this Section we have compared the performances of $l1$-logit and our hybrid approach during the analysis of the SWPC test set covering the time range between August 1996 and December 2010 (the cardinality of such set is $22222$); for both methods we used the data collected between December 1988 and June 1996 as training set (the cardinality of this second set is $17600$). Further, we have also analyzed the same test set by means of other four classical machine learning methods: the (unsupervised) clustering HCM and FCM algorithms, a standard Multi Layer Perceptron (MLP) \citep{ruetal86} and a Support Vector Machine (SVM) \citep{cortes_support-vector_1995}. For the latest two methods, which are supervised, we used the same training set as in the case of $l1$-logit and the hybrid method. All these prediction algorithms have been applied to predict flares with class above $C1$ and $M1$, respectively. From now on, for sake of brevity, we will indicate with $\geq C1$ and $\geq M1$ all flares with class above $C1$ and $M1$, respectively. We have not considered flares with class above $X1$ since they are rare in this dataset (less than $1\%$ in the training set and around $0.5 \%$ in the test set).

By means of the frequency matching process described in Section 2, each sample is transformed into a $5$-dimensional vector. Note that the first four components range from $0$ to $1$, while the fifth one, i.e., the sunspot area, goes from $0$ up to $10^2$. Since the differences between component variances can affect the flare prediction performances, a standardization step preceded the application of the machine learning algorithms. We also note that frequency matching must be performed for each case of interest, i.e, separately for the $\ge C1$- and $\ge M1$-flare predictions; therefore, for both the training set and the test set, we have constructed two subsets: the first subset, indicated with $\#1$, is constructed using the frequencies of flares of class at least C1 (i.e., by applying (\ref{eq:fc1}) and analogous); the second subset, indicated with $\#2$, to the frequencies of flares of class at least M1 (i.e., by applying (\ref{eq:fm1}) and analogous).

\begin{figure}[ht!]
\subfigure[]
{
\includegraphics[width=0.49\textwidth]{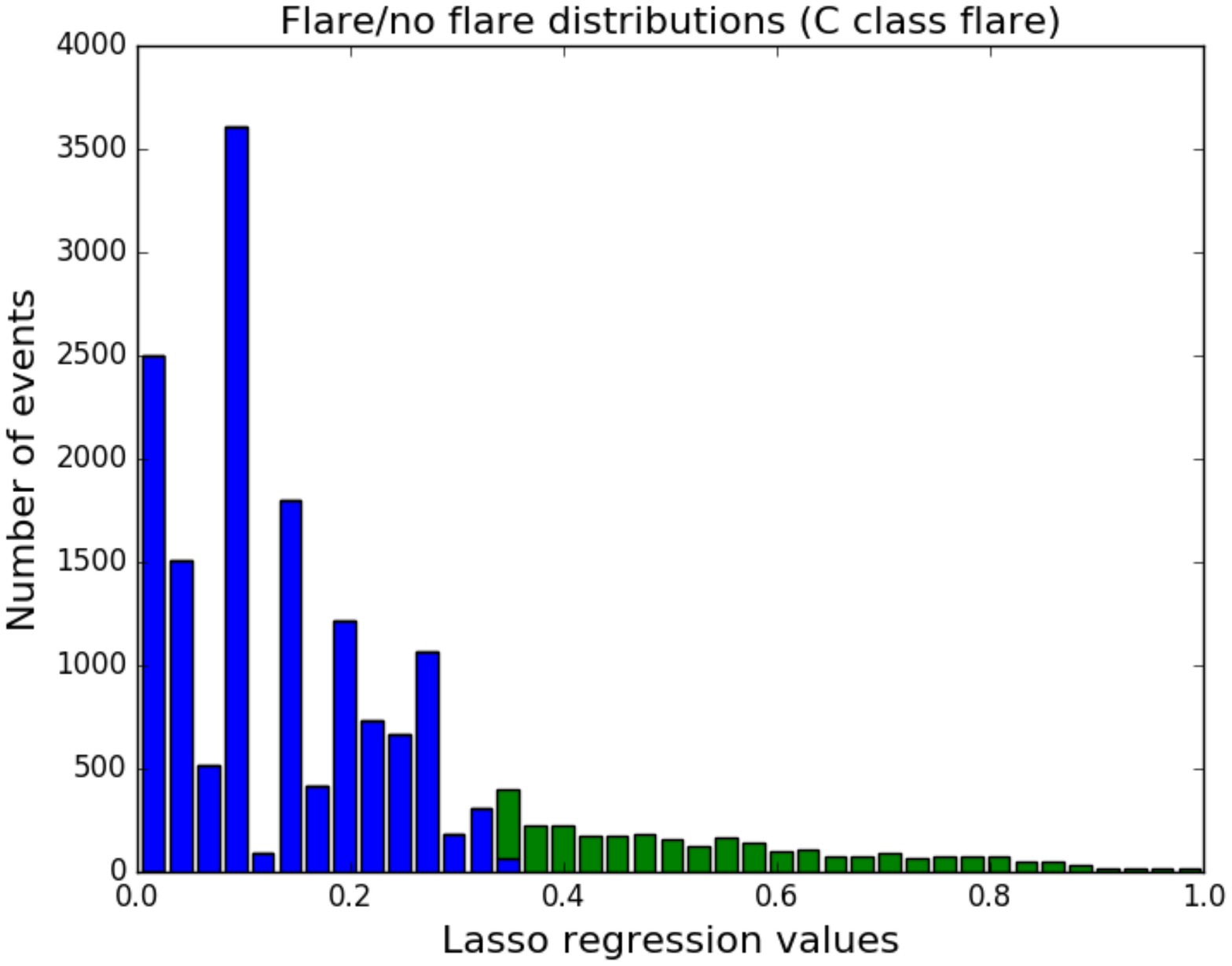}
}
\subfigure[]
{
\includegraphics[width=0.49\textwidth]{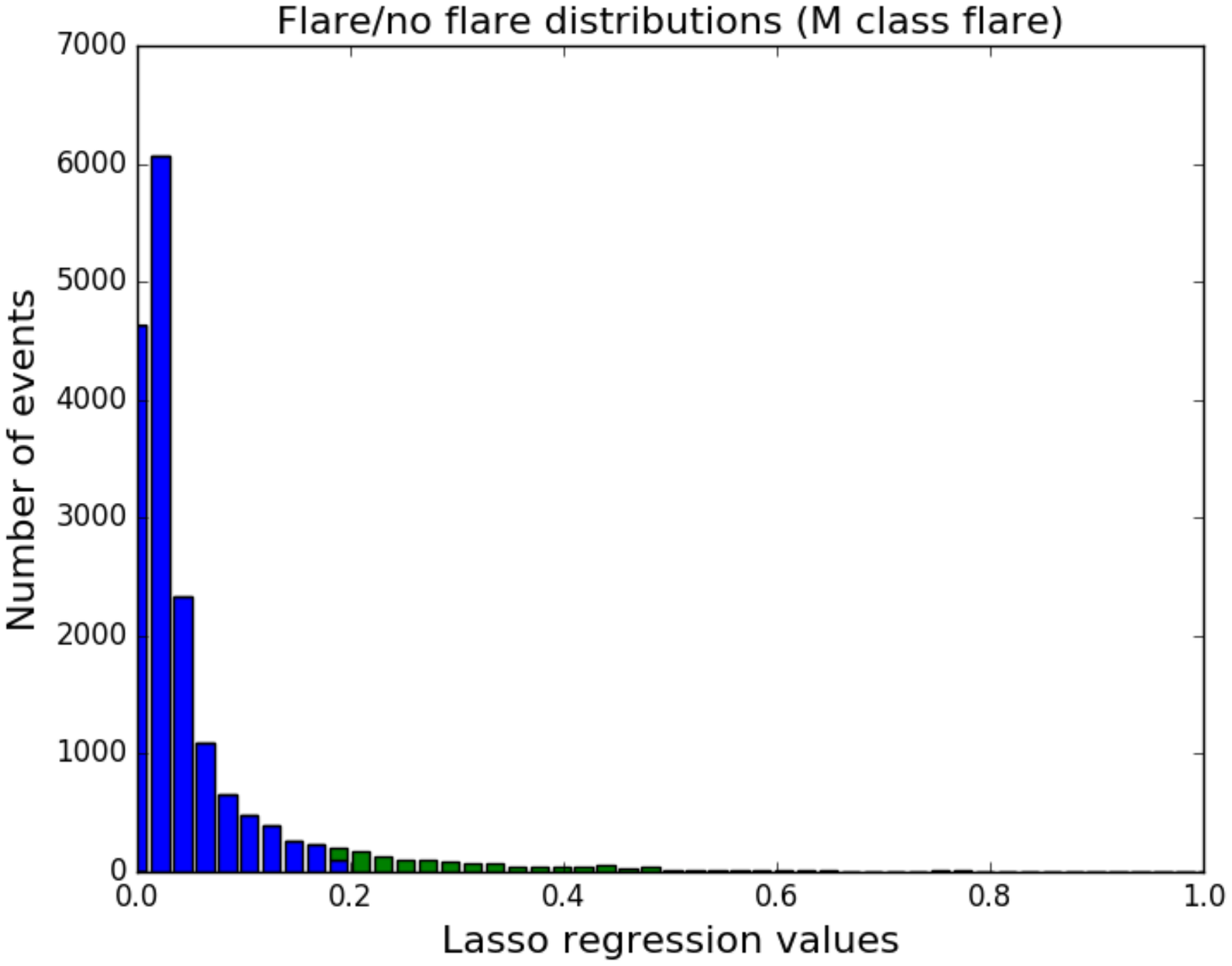}
}
\caption{(a) $\geq C$ class flare prediction. Split of the Lasso regression output by means of the Fuzzy C-means algorithm.
The x-axis shows the values of the regression outcomes provided by the cross validated Lasso algorithm.
Blue and green colors represent the two clusters identified by the Fuzzy C-means algorithm.
Blue (resp. green) cluster is the set of all the events for which the hybrid method returns a no-flare (resp. flare) prediction. (b) The same as in (a) but for $\geq M$ class flares.}
\label{fig:fig0} 
\end{figure}

As explained in the previous section, the main advantage of the hybrid approach is in the fact that the way it partitions the set of LASSO outcomes is driven by the input data. This is clearly described in Figure \ref{fig:fig0}, showing how FCM automatically identifies the probability threshold. It is interesting to note that this threshold depends on the flare class under consideration and in any case is different than the fixed value provided by $l1$-logit, which splits the regressions values at $0.5$.

\begin{figure}[ht!]
\plotone{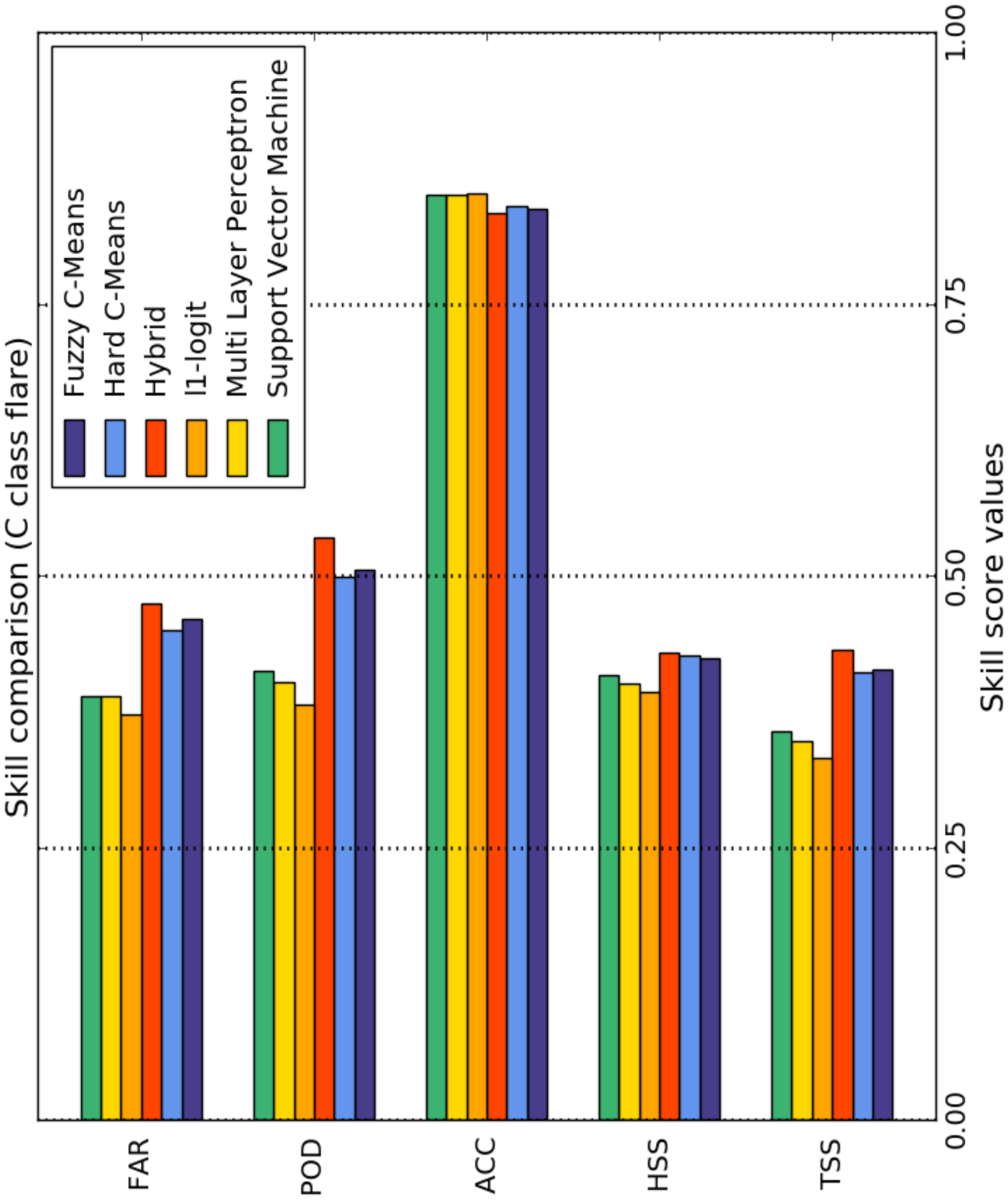}
\caption{Comparison of performance between the six flare prediction algorithms in terms of skill scores. 
The bar plots represent the skill score values obtained by applying each method to the test set for the prediction of $ \geq C1$ flares.}
\label{fig:f1} 
\end{figure}

\begin{figure}[ht!]
\plotone{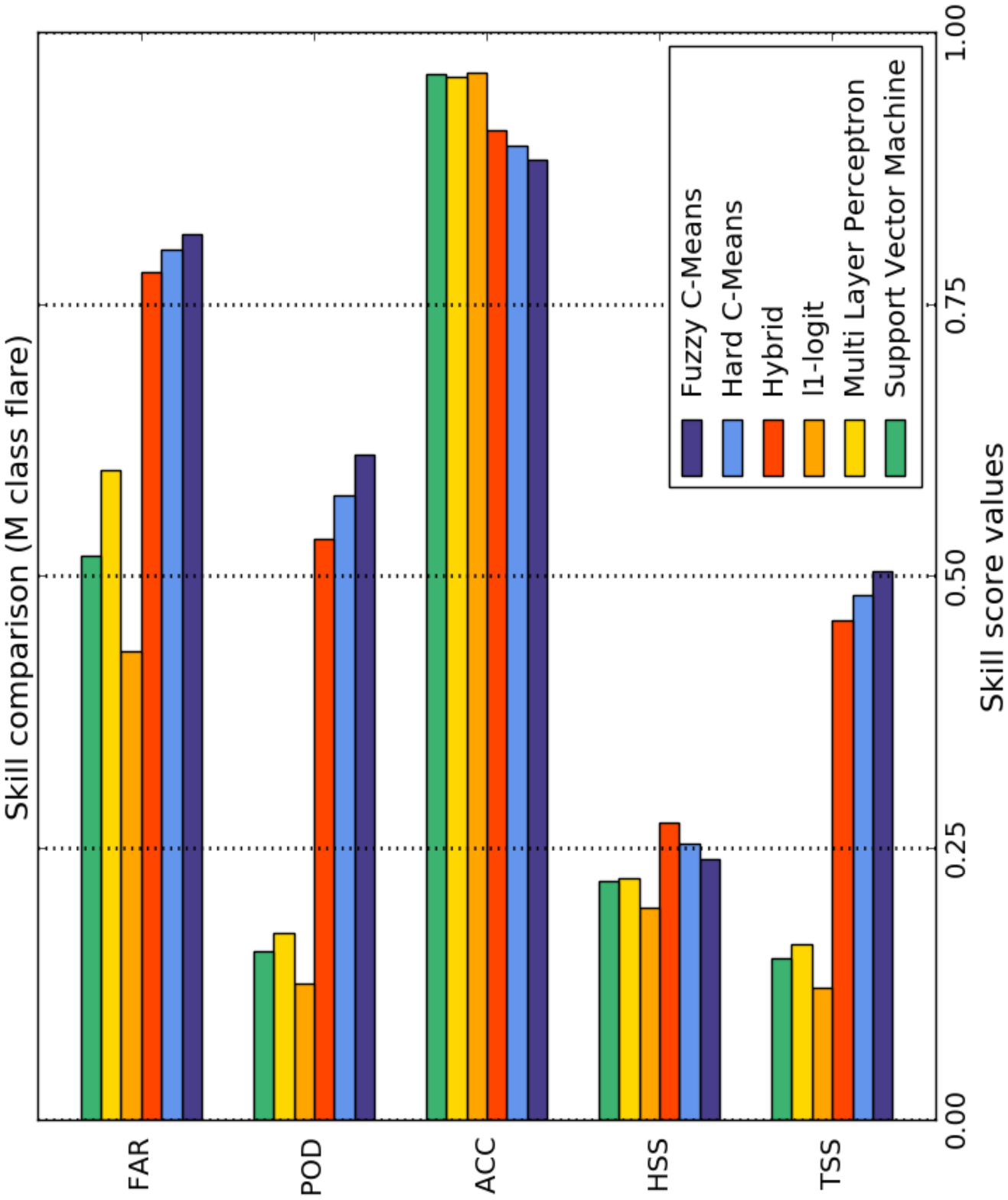}
\caption{The same as in Figure \ref{fig:f1} but for the prediction of $ \geq M1$ flares.}
\label{fig:f2} 
\end{figure}

The threshold value determines the prediction, whose performance can be measured by means of specific scores. Many skill scores can be found in literature for the assessment of flare prediction performances \citep{bletal12}. All these scores are linked to the forecast contingency tables made up of four elements:
\begin{itemize}
\item The number of flares predicted and observed (true positives, TP).
\item The number of flares not predicted but observed (false negatives, FN).
\item The number of flares predicted but not observed (false positives, FP).
\item The number of flares not predicted and not observed (true negatives, TN).
\end{itemize}
We have validated the six flare prediction algorithms by means of the following skill scores defined in terms of the above elements. Specifically, the probability of detection
\begin{equation}\label{pod}
POD = \frac{TP}{TP+FN}~;
\end{equation}
the accuracy
\begin{equation}\label{acc}
ACC = \frac{TP + TN}{TP + TN + FP + FN}~;
\end{equation}
the false alarm ratio
\begin{equation}\label{far}
FAR = \frac{FP}{TP + FP}~.
\end{equation}
These scores range from $0$ to $1$ and best predictions correspond to small FAR values and high values for the other scores. We also utilized two scores with values ranging from $-1$ to $1$: the Heidke skill score
\begin{equation}\label{hss}
HSS = \frac{2 \cdot (TP \cdot TN - FN \cdot FP)}{(TP + FN) \cdot (FN + TN) + (TP + FP) \cdot (FP + TN)}~;
\end{equation}
and the true skill statistics
\begin{equation}\label{tss}
TSS = \frac{TP}{TP+FN} - \frac{FP}{FP + TN}~.
\end{equation}
Also in this case good prediction performances correspond to high values of the scores. Figure \ref{fig:f1}  and Figure \ref{fig:f2} present the values of all five skill scores for the $\geq C1$ flare prediction and the $\geq M1$ flare prediction, respectively. Moreover, Table 1 and Table 2 provide the results of the feature selection processes preformed by $l1$-logit and the hybrid technique. Specifically, the tables contain the weights $\beta$ with which the sunspot area, the McIntosh indices, and the Mount Wilson index contribute to the flare prediction process for the two methods.

\startlongtable
\begin{deluxetable}{cccccc}
\tablecaption{Feature importance in $\geq C$ class flare prediction computed from the training set. For each method, the values correspond to the weights associated to the features divided by the sum of the weights, i.e. $\bar \beta_k := \hat \beta_k / \sum_{j=1}^F \hat \beta_j $.}
\label{tab:C-flare}
\tablehead{
& \colhead{MtWilson} & \colhead{McIntosh $Z$} & \colhead{McIntosh $p$} & \colhead{McIntosh $c$} & \colhead{Area} \\
}
\startdata
hybrid & $0.198$ & $0.268$ & $0.222$ & $0.164$ & $0.147$ \\
$l1$-logit & $0.189$ & $0.332$ & $0.219$ & $0.154$ & $0.104$
\enddata
\end{deluxetable}

\startlongtable
\begin{deluxetable}{cccccc}
\tablecaption{Feature importance in $\geq M$ class flare prediction computed from the training set. See caption of Table 1 for the meaning of table entries.}
\label{tab:M-flare}
\tablehead{
& \colhead{MtWilson} & \colhead{McIntosh $Z$} & \colhead{McIntosh $p$} & \colhead{McIntosh $c$} & \colhead{Area} \\
}
\startdata
hybrid & $0.281$ & $0.117$ & $0.047$ & $0.146$ & $0.407$ \\
$l1$-logit & $0.181$ & $0.264$ & $0.217$ & $0.142$ & $0.194$
\enddata
\end{deluxetable}

\section{Discussion and conclusions}

This paper introduces a novel approach to flare prediction, which utilizes indices associated to ARs data and which is also able to automatically indicate the ones, among such features, that mostly contribute to prediction. The approach is intrinsically hybrid, in the sense that it is based on the combination of the ability of regularization to perform feature selection with the ability of clustering to classify in a data-adaptive fashion. In the present implementation we have used LASSO in the feature selection step and FCM in the clustering step. In fact, LASSO guarantees a notable degree of generality in regularization while FCM guarantees a notable degree of flexibility in data adaptation. Anyhow, we have tested the hybrid approach using different combinations of feature selection and clustering methods involving $l1$-logit and HCM: the results of both feature importance ranking and prediction were comparable.

We validated the approach against a NOAA SWPC dataset and by comparing the results with the ones provided by $l1$-logit and other standard machine learning flare prediction algorithms. This comparison showed that the hybrid approach outperforms $l1$-logit in the case of HSS and TSS, that are often considered \citep{bletal12} the most reliable skill scores in the game (for example, ACC tends to reach its maximum when the threshold is $0.5$, which is not fully appropriate in the case of unfrequent events such as M and X class flares). This is particularly true for TSS and for the prediction of flares belonging to class M or higher. More in general, the hybrid method predicts with a performance rate which is very similar to the one of the other two unsupervised clustering algorithms, while, coherently, $l1$-logit works similarly to the other two supervised regularization methods. The higher forecasting effectiveness of the hybrid approach with respect to $l1$-logit is due to the fact that it performs classification with a thresholding procedure which is data adaptive, while $l1$-logit utilizes a fixed negative/positive threshold. We note that the threshold in $l1$-logit could be tuned heuristically, searching `a posteriori' for the values that provide the maximum for TSS and HSS and that the advantage of fuzzy clustering is that it realizes such search `a priori' and in an automatic way.

The hybrid approach and $l1$-logit can be compared also as far as their feature selection power is concerned. Table 1 clearly shows that, in forecasting $\geq$ C1 flares, the two methods indicate the same features as the ones that mostly contribute to the prediction. Results are different when predicting $\geq$ M1 flares, since LASSO gives the highest emphasis to the AR area, while $l1$-logit points out more significantly two of the three McIntosh indices as mostly significant. A clarification of this contradictory outcome shall be obtained by means of a systematic application of these two methods against either several SWPC datasets or features extracted from SDO/HMI images; this activity is part of the tasks currently addressed by the H2020 project FLARECAST, which will provide a technological platform for the testing of flare prediction algorithms and for the validation of the forecasting and feature selection results.

\acknowledgments
The authors have been supported by the H2020 grant Flare Likelihood And Region Eruption foreCASTing (FLARECAST), project number 640216. The authors kindly thank Prof. Shaun Bloomfield for providing the SWPC data and Dr. Annalisa Perasso for useful discussion. 

\bibliography{swpc}

%
%



\end{document}